# Characterizations of strain and defect free GaN nanorods on Si(111) substrates

H. W. Seo[a], Q. Y. Chen[b], M. N. Iliev, and W. K. Chu

*Department of Physics and Texas Center for Superconductivity & Advanced Materials, University of Houston, Texas*

L. W. Tu, and C. L. Hsiao

*Department of Physics and Center for Nanoscience & Nanotechnology*

*National Sun Yat-Sen University, Taiwan, Republic of China*

James K. Meen

*Department of Chemistry and Texas Center for Superconductivity & Advanced Materials, University of Houston, Texas*

## Abstract

GaN-nanorods grown on Si(111) substrates are found strain- and defect-free as characterized by micro Raman spectroscopy, secondary electron (SE) and cathode-luminescence (CL) imaging. The matrix supporting the nanorods bears the brunt of all strains, strain-relaxations, and defect generations, giving the nanorods an ideal environment to grow to perfection. Photo-excitations by the Raman laser source and electron irradiation during CL imaging lead to an increase of non-equilibrium electrons, suggesting an effective approach to photo-emitting or field emitting device applications. The nanorods, largely isolated from but perfectly aligned with the sustaining matrix, are grown in excellent epitaxy with the Si substrates.

a) E-mail: hseo@uh.edu, b) qchen@uh.edu, corresponding author.

Wurtzite GaN has been a subject of intensive studies in the past few years due to its unique physical properties such as high melting point, high breakdown voltage, large bandgap, stable chemical properties, and a hardness equivalent to that of diamond. It can be easily doped with Si to become an n-type semiconductor that is suitable for use as a field emitter. More importantly, its direct bandgap, which can be engineered by alloying with In and Al to cover most of the visible spectral range, makes it one of the most intensively sought-after color-generating materials.[1-3] In addition, as quantum effects emerge when the material is fabricated into nanostructures, one can envisage many new nano-device application possibilities.[4-6] Among various nanostructures, nanorod is particularly interesting for its crystallographic alignment to many industrial substrate materials and high length-to-diameter aspect ratio. Its existence in orderly array, furthermore, is in sharp contrast to the entangled nanowire structures observed in materials grown with metal-particle seeds.[4] While the film growth and growth mechanism of GaN nanorods have been reported before, [7, 8] in this paper we will focus mainly on the crystallographical and microstructural aspects. Microstructural imperfections and strain states of the nanorods were studied at length with respect to the matrix areas because structural defects can greatly limit device performance and have been among the most common reasons why a device fails.

The samples studied in this work were grown by nitrogen plasma enhanced molecular beam epitaxy (MBE), as reported earlier in ref. 7. Shown in Fig. 1, the nanorods grow as standing posts protruding from craters. The lateral dimensions range from ~10-100 nm and longitudinal dimensions from ~0.6-0.8 μm above the film surface. These nanorods spread around various parts of the sample, but sometimes congregate into bundles as shown in the inset. The full lengths of the nanorods are ~2.4 -2.6 μm, giving an aspect ratio of 20-250. From a cross-sectional view, the nanorods are found embedded in nanotrenches, and the pin-shape bottom serves as a point-like support to the nanorods.[7] Therefore, in comparison with the matrix areas, the nanorods are expected to experience different lattice-mismatch induced stress fields during film growth as well as different thermal stress build-ups upon cooling. Since the resulted strain and its subsequent relaxation process are known to be a main driving force for the formation of vacancies, interstitial defects, and dislocations, it is understandable that the structural imperfections of the nanorods can be quite distinct from those of the matrix. More specifically, it is natural to expect the nanorods to be less imperfect since they are in some way isolated from the bulk of the strain field and the dynamics of strain relaxation. To verify this argument, we first present in what follows the Raman spectroscopic characterizations on the strain field and carrier concentration, then show the spatially-resolved cathodoluminescence (CL) imaging and Electron Backscatter Diffraction (EBSD) results obtained using a scanning electron microscope over various parts of the samples.

Micro-Raman spectrometry was performed using the 632.8 nm line of a He-Ne laser focused through an x100 objective into a spot of 1-2 μm in diameter. For this objective lens and laser wavelength, the depth of focus is estimated to be ~2 μm and, therefore, the Raman signals emitted from the 1.8 μm thick GaN film are inevitably superimposed with those from Si substrate near the GaN/Si interface. The Raman spectra were collected separately from both the nanorod-clustered areas and the matrix area (used herein as a control) with $z(xx)\underline{z}$ backward scattering configuration, where $z$ // [0001] and $\underline{z}$ // [000$\underline{1}$] are, respectively, the directions of incident and scattered light while $xx$ denotes the polarizations of the incident (x) and scattered lights (also x), here with both *x*'s // $[110]_{Si}$ // $[2110]_{GaN}$. This correlation of crystallographic orientations will become clear when the EBSD data are presented.

The nanorod-abundant regions are indeed largely strain-free as compared to the matrix region of no nanorods. Indeed, the Raman data of Fig.2 show that the GaN $E_2$ (high) mode is at 567.7 $cm^{-1}$ for the nanorod-abundant area as compared to 566 $cm^{-1}$ for the matrix, both in reference to the bulk material's 567.7 ± 0.5 $cm^{-1}$.[9-12] The 1.7 $cm^{-1}$ red shift in the matrix areas suggests that the matrix is still under tensile stress by about 0.5 GPa throughout all the film thickness.[9-12] This strain state can be either from remnant (viz. incompletely relaxed) lattice mismatch stress or from residual thermal stress. The 566 $cm^{-1}$ peak, we observe, contains a component of the strain-free 567.7 $cm^{-1}$ peak, as judged by its broadness. This being truly the case, then the matrix area must comprise a layer of tensile strain (the buffer layer) and a layer which is strain-free (the upper layer of the film). Another interesting observation is on the existence or absence of the $A_1$ (LO) mode at 733-736 $cm^{-1}$, as shown in the inset of Fig. 2. This vibration mode is observed in highly electrically resistive GaN materials with a carrier concentration up to $10^{18}$ $cm^{-3}$, beyond which the mode is absent.[2, 13] The broadness of this $A_1$ (LO) peak is suggestive, again, of the tensile-stress in the buffer layer (733 $cm^{-1}$) and the strain-free nature of the upper layer (736 $cm^{-1}$).[2]

The disappearance of $A_1$ mode in the nanorod-populated area is indeed conspicuous, implying higher electron concentration (*n*) compared to the matrix area. Such higher *n* can tentatively be explained by geometric effects, because photo-induced electric charges can build up around the tip of a nanorod. The enhancement factor ($\beta$) has been shown, to first order, to scale linearly with the aspect ratio ($\eta$) by $\beta \approx 2+\eta$,[14-16] except that, instead of an externally applied voltage, in the present case the higher charge carriers are established mainly by photo-excitation via the Raman light source. But subsequent shifts in quasi-Fermi level due to the photoelectron generation give rise to an equivalent electric potential difference. In this spirit, excitation by electron irradiation observed in the CL mode should also work to the same effect.

The CL images were obtained using a 15 keV electron beam, where signals were collected from the top ~1.3 μm thickness of the sample at room temperature. The broad spectrum of yellow luminescence is known to arise from the existence of Ga vacancy ($V_{Ga}$) in an n-type GaN material because $V_{Ga}$ introduces deep accepter levels at 0.8~1 eV above the valence band edge, which provide the necessary passages for donor-accepter optical transitions.[1, 17, 18] With this understanding, the spatially resolved CL images (Fig. 3 (a)) of the nanorod (dark area) and nanotrench (bright area) in the visible spectral range then suggest that the nanorods should contain no obvious defect states, but what surrounds the nanotrench is more defective. The existence of $V_{Ga}$ compensates the shallow donors, resulting in a reduced *n*, as is reflected in the non-diminishing $A_1$ mode even under laser illumination.[19] This is consistent with our conclusion that the higher *n* in the nanorod region inferred from the Raman data is mainly due to geometric effects, namely the high aspect ratio of the rods, rather than from the defects. Our earlier PL result does support this argument because the nanorod areas had shown only excitonic states.[7] From the inhomogeneous grey level contrast of the CL images, we envision that the matrix area contains a large amount of non-uniformly distributed structural imperfections. All told, the defective matrix areas must have borne the brunt of all strain generation and relaxation, giving the simultaneously growing nanorods an essentially strain-free environment. To further elucidate our point, we show in Fig. 3(b) the conjugate images of the matrix in CL and SEI modes; here we note that the CL image is largely a negative to the secondary electron image (SEI), which further supports our reasoning of structural-defect induced luminescence.[20-22]

Despite the ubiquitous point defects throughout the matrix regions, the overall crystallinity and epitaxial quality of the samples are still excellent. We have studied the crystallographic orientations and epitaxial quality by EBSD measurement (Fig 4) based on Kikuchi pattern recognition. In order to maximize the Kikuchi pattern formation, the 20 keV incident electron beam, with a nominal spot size set at ~50 nm, was tilted 70° away from the sample normal direction. The apparent spot size of scanning beam, we note, could have been much larger than the intended 50 nm, say ~100 nm, because of instrument vibration or scanning beam instability. This larger beam spot size brings about certain negative consequences on the spatial resolution, though this electron beam technique still remains to be one of the most unique approaches among all other spatially resolved analytical methods in nano-technology due to its convenient beam spot-size controlling and short data acquisition time capabilities.

The EBSD data in Fig 4 show that the nanorods have excellent epitaxial quality with rod axis // <0001> of the matrix area. In addition, the close-packed <2110> axis for both the nanorods, nanotrenches and matrix regions are all parallel to the <110> close-packed direction of the {111} oriented cubic Si. This is very interesting considering that the nanorods are physically separated from the matrix volume except at the pin-shaped bottom.

In summary, Raman spectroscopy, SEI, CL, and EBSD confirm that wurtzite GaN nanorods grow epitaxially on Si(111) substrates and are strain-free with buffer layers bearing much of the remnant stress. The outstanding structural integrity and perfection accompanied with the high aspect ratio of 20-250 may find useful applications in field emission electron devices. The reduced amount of defects in the nanorod area also is an important feature for practical applications because structural imperfections cause yellow luminescence and are among the most common reasons why electronic or optoelectronic devices fail. In addition, well aligned crystallographic orientation of nanorod could also be exploited through the anisotropy of nanostructures in luminescence, refraction, and dielectric functions, among many other physical properties.[23]

This work was supported by the State of Texas through the Texas Center for Superconductivity and Advanced Materials at the University of Houston and in part by the Welch Foundation. Work at National Sun Yat-Sen University was supported by the National Science Council of the Republic of China.

Figure Captions

Fig. 1

SEM image (SEI) of GaN nanorods, showing posts protruding out of craters. Inset: nanorod clusters (20 º tilt view), Scale bar :100 nm.

Fig. 2

Raman shifts showing the strained and strain-free states of GaN nanorods, GaN matrix materials and the silicon substrate. The 1.7 $cm^{-1}$ shift with respect to the bulk value represents 0.5 GPa of thermal stress in the matrix. The nanorod region shows no shift and hence is strain-free. Inset : the $A_1$ (LO) mode of the matrix and nanorod regions.

Fig. 3 (a)

Top view SEM images showing nanorod (center), encompassing nanotrench, and the surrounding matrix. Left panel: SEI mode, right panel: CL imaging (CLI) mode. Scale bar: 100 nm. In CLI mode, dark area manifests defect-free region in contrast to the bright, defect-containing regions, contrary to the SEI mode.

Fig. 3 (b)

Conjugate images of a matrix region. Left panel: SEI mode, right: CLI mode, scale bar: 100 nm. Corresponding positions are labelled as (a, b, c, d) vs. (a´, b´, c´, d´) to help depict the negative contrast (see part (a) for descriptions on darkness and brightness.)

Fig. 4

EBSD Kikuchi patterns of a nanorod showing the relative orientations of a (0001) GaN film on a Si(111) substrate, showing excellent crystalline diffraction patterns of both the nanorods and the matrix material as well as clear epitaxy and crystallographic alignment. (a) nanorod-abundant regions, (b) matrix regions, and (c) bare Si substrate areas stripped of the film.

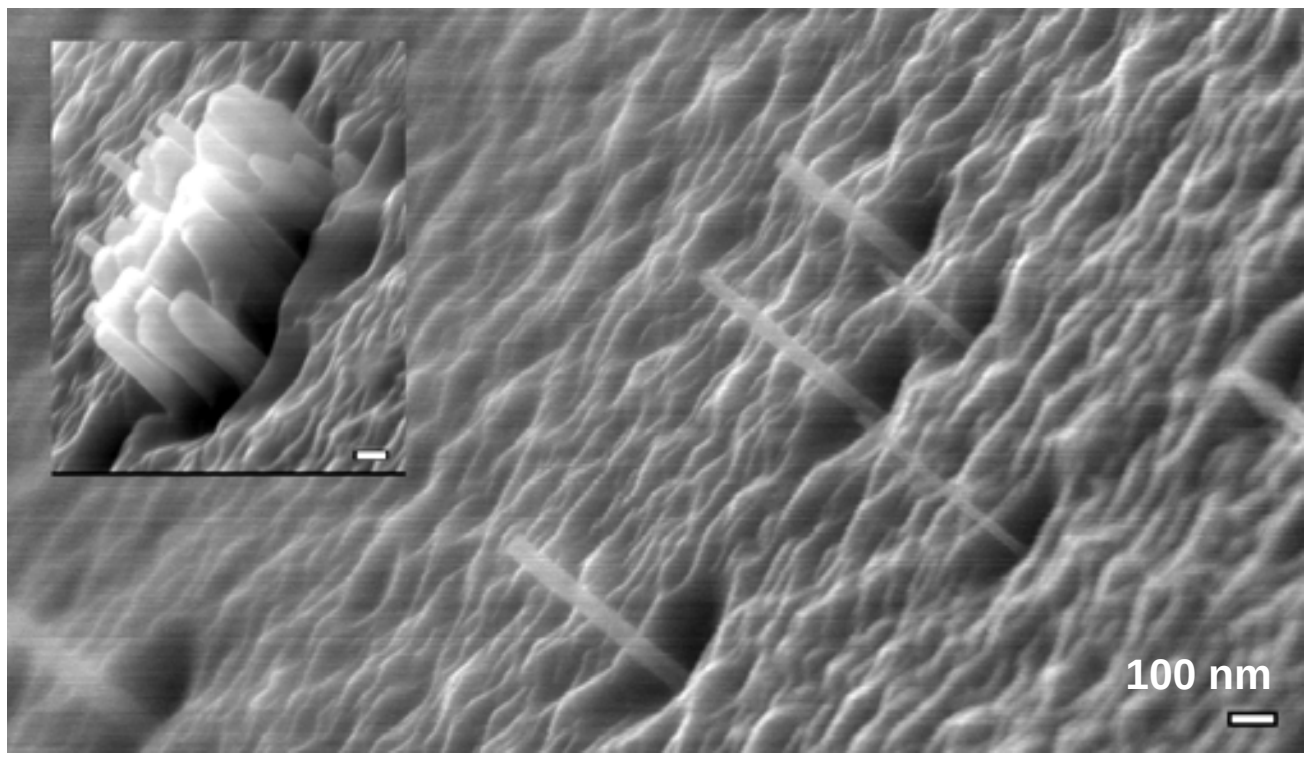

[Fig. 1]

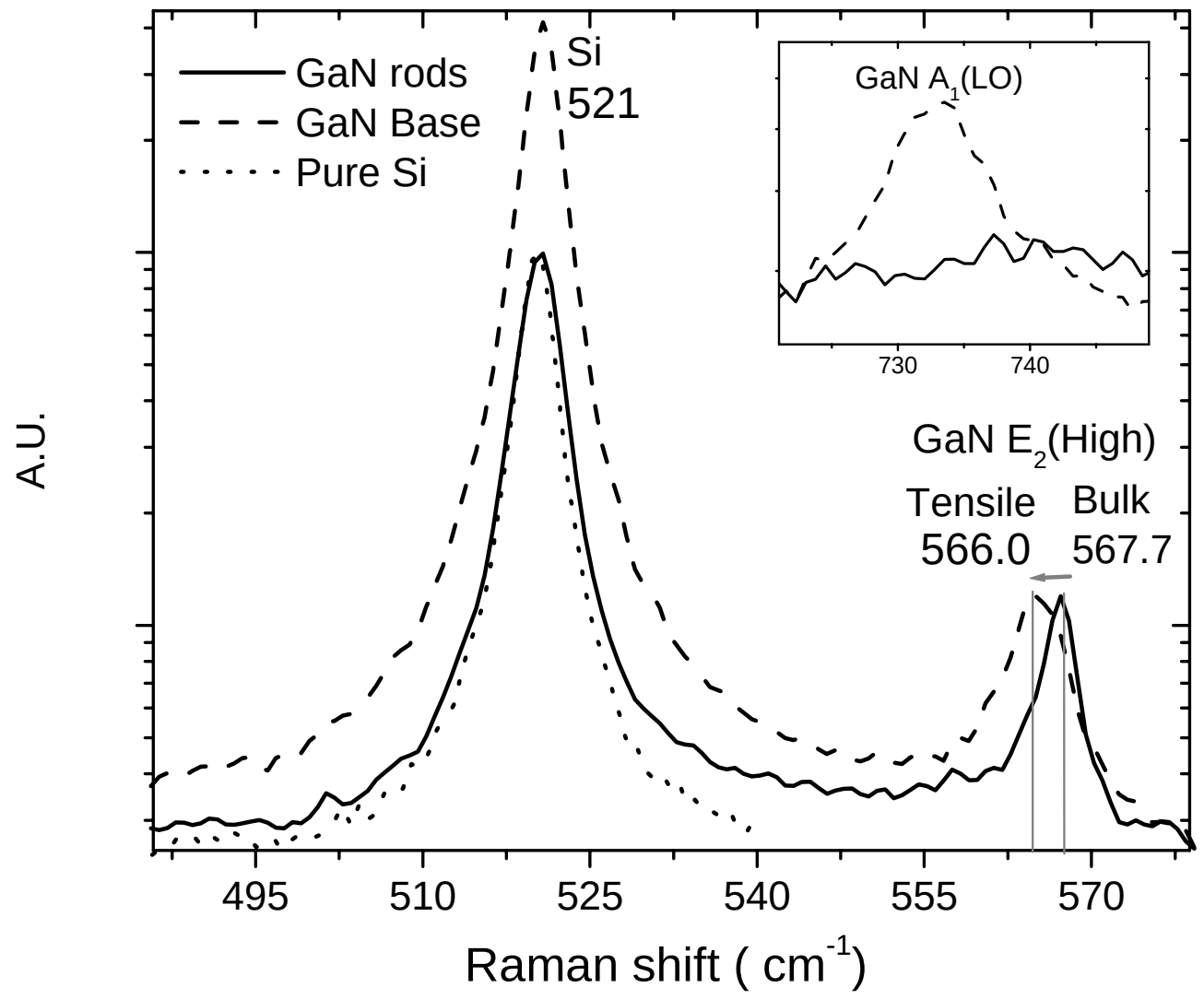

[Fig. 2]

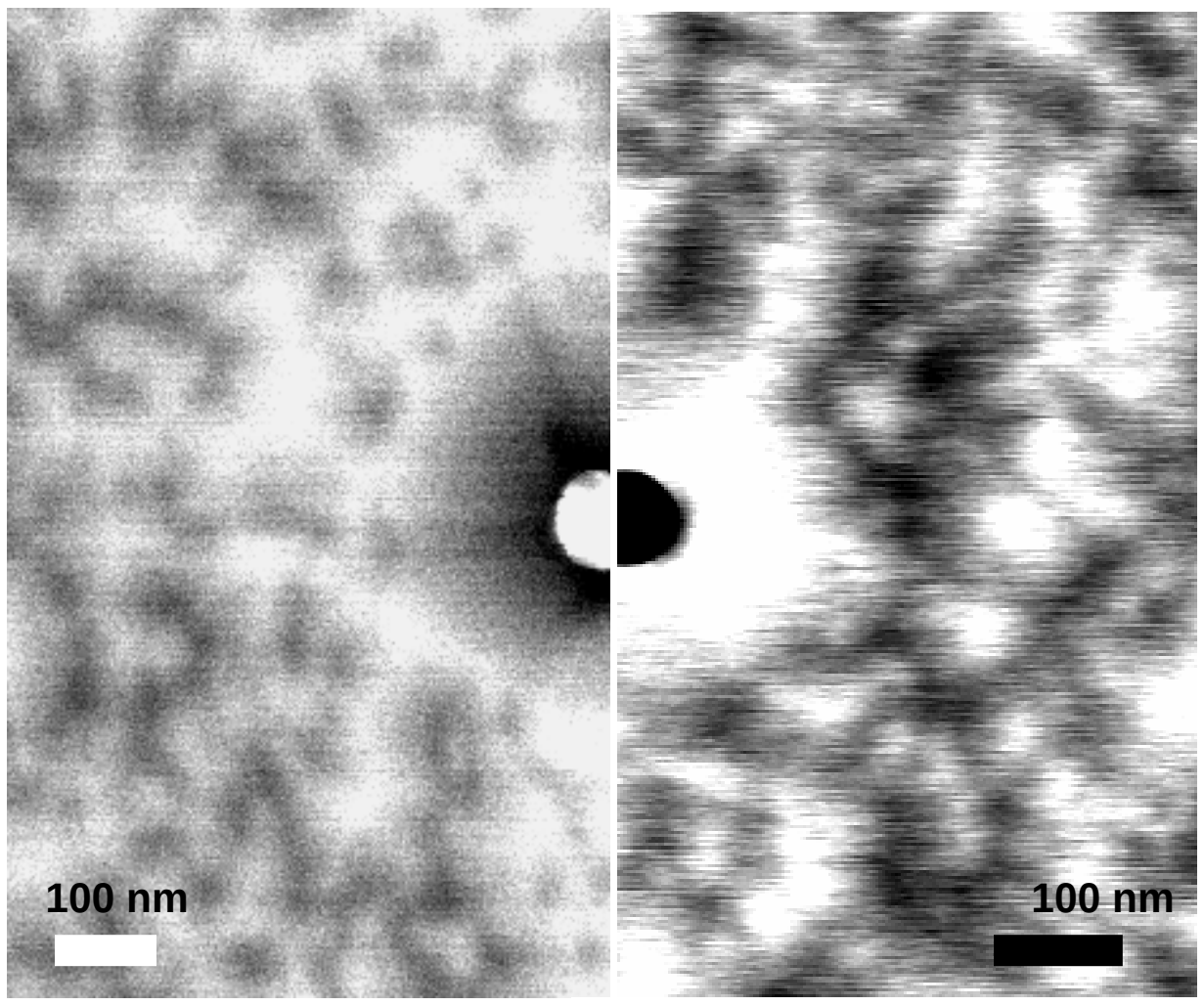

[Fig. 3 (a)]

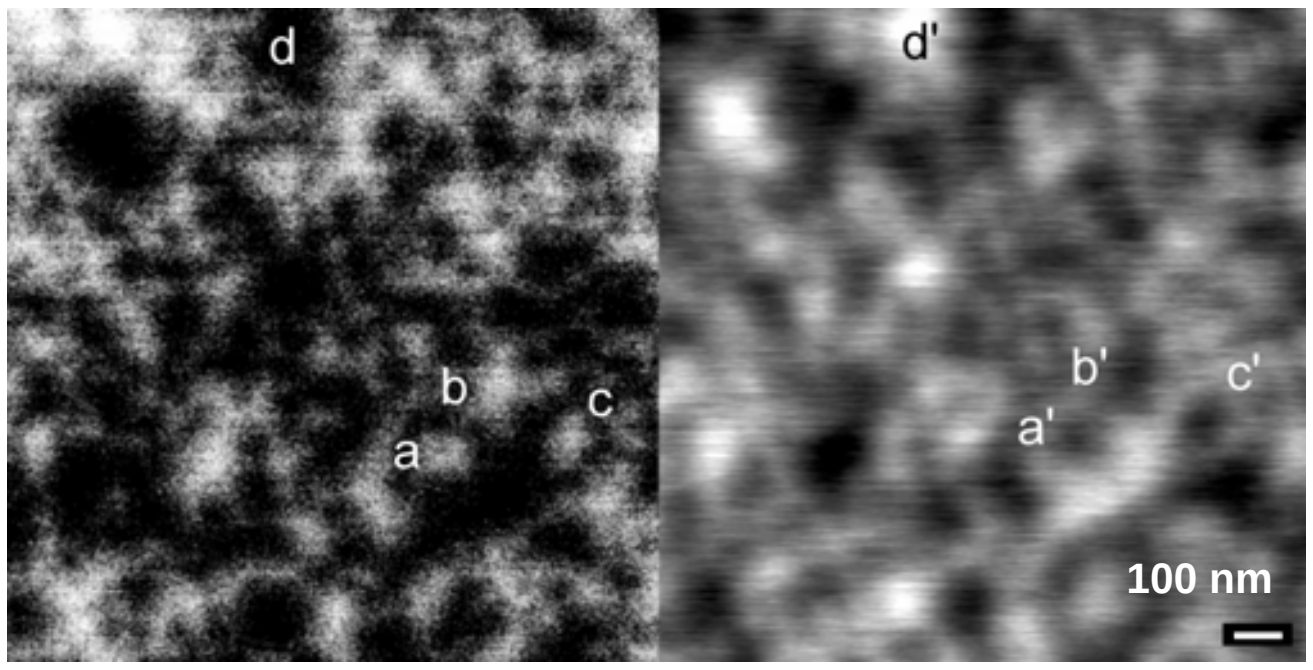

[Fig. 3 (b)]

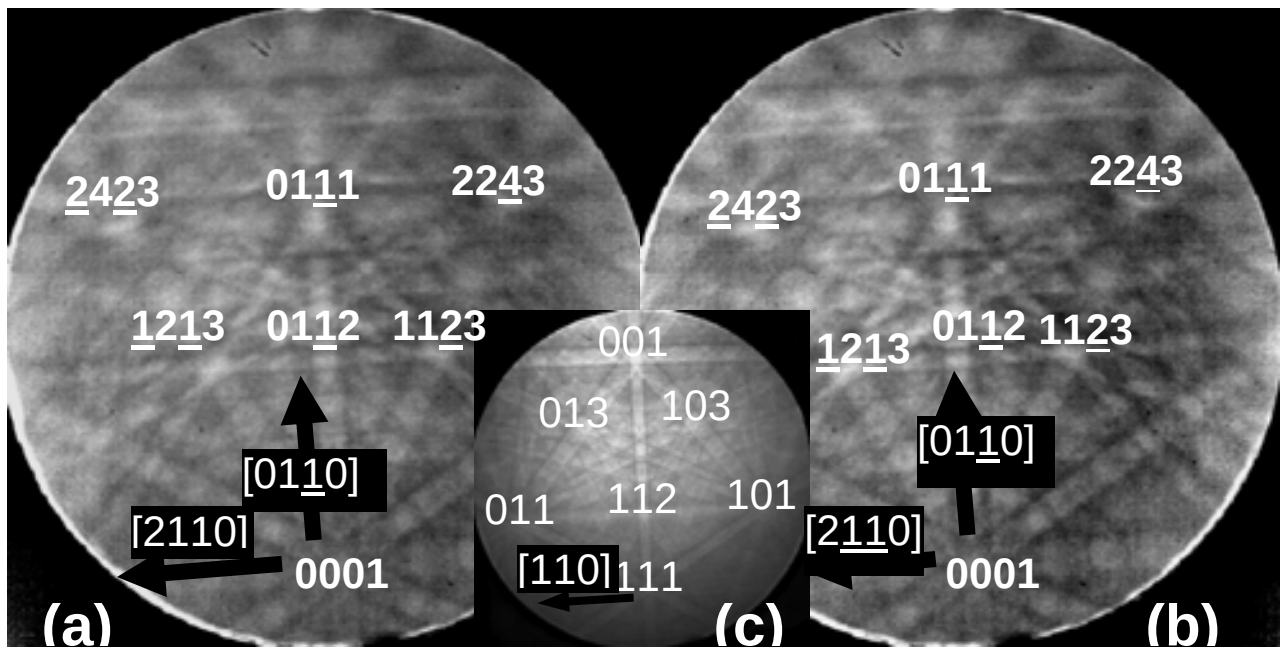

[Fig. 4]